\documentclass[aps,prb,twocolumn,groupedaddress]{revtex4}

\usepackage[pdftex]{graphicx}
\usepackage{amsmath}
\usepackage{amssymb}
\def \etal {{\it et al.\ }}

\begin{document}

%\preprint{}

\title{Three-dimensional Dirac fermions in quasicrystals as seen via optical conductivity}

\author{T. Timusk and J.P. Carbotte}
%\thanks{}
\affiliation{Department of Physics and Astronomy, McMaster University, Hamilton, ON L8S 4M1, Canada}
\affiliation{The Canadian Institute of Advanced Research, Toronto, Ontario M5G 1Z8, Canada}

\author{C.C. Homes}
\affiliation{Condensed Matter Physics and Materials Science Department, Brookhaven National Laboratory, Upton, New York 11973, USA}

\author{D.N. Basov}
\affiliation{Department of Physics, University of California at San Diego, La Jolla, California 92093-0319, USA}

\author{S.G. Sharapov}
\affiliation{Bogolyubov Institute for Theoretical Physics, National Academy of Science of Ukraine, 14-b Metrologicheskaya Street, Kiev, 03680, Ukraine}
\affiliation{Department of Physics, Taras Shevchenko National Kiev University, 6 Academician Glushkov Ave.,
Kiev 03680, Ukraine}

\date{\today}

\begin{abstract}
The optical conductivity of quasicrystals is characterized by two features not seen in ordinary metallic systems.  There is an absence of the Drude peak and the interband conductivity rises linearly from a very low value up to normal metallic levels over a wide range of frequencies. The absence of a Drude peak has been attributed to a pseudogap at the Fermi surface but a detailed explanation of the linear behavior has not been found.  Here we show that the linear conductivity, which seems to be universal in all Al based icosahedral quasicrystal families, as well as their periodic approximants, follows from a simple model that assumes that the entire Fermi surface is gapped except at  a finite set of Dirac points.  There is no evidence of a semiconducting gap in any of the materials suggesting that the Dirac spectrum is massless, protected by topology leading to a Weyl semimetal.  This model gives rise to a linear conductivity with only one parameter, the Fermi velocity.  This picture suggests that decagonal quasicrystals should, like graphene, have a frequency independent conductivity, without a Drude peak.  This is in accord with the experimental data as well.

\end{abstract}

\pacs{}
%\keywords{}

\maketitle

\section{ Introduction}
The electronic properties of quasicrystals\cite{shechtman84} are not what one expects of alloys of good metals.  The more perfect the crystals are, the worse is their electrical conductivity, just the opposite  of what one sees in the transport properties of their metallic constituents where the residual resistance decreases as order and purity are improved\cite{poon92}.  The best quasicrystals  are almost insulators with a resistivity as high as 30 m$\Omega$cm and a semiconducting temperature dependence\cite{biggs90}.  The high resistivities are not caused by excessive disorder as seen in metallic glasses\cite{theye85}. High quality decagonal quasicrystals, have a quasicrystalline structure in two dimensions but are ordinarily periodic crystals in the third direction.  Their electrical conductivity is low and frequency independent in the quasicrystalline plane, but in the periodic c-direction the decagonal quasicrystals behave like ordinary metals: there is a Drude peak and the conductivity is high with a metallic temperature dependence\cite{shuyuan90,martin91}.

The generally accepted model for this near-insulating behavior of quasicrystals is a pseudogap that destroys a large portion of the Fermi surface leaving only small pockets of electrons that can contribute to conduction.\cite{pierce93}  The low electronic specific heat coefficient $\gamma$ is consistent with this picture\cite{biggs90}.  Recent  HAXPEX spectra show a clear pseudogap in the density of states of  Al$_{63}$Cu$_{25}$Fe$_{12}$ at the Fermi surface\cite{nayak12}. The  origin of this pseudogap can be understood in terms of the Hume-Rothery rules for the formation of stable alloys\cite{bancel86,smith87,trambly95}. The overall energy of the system can be reduced if the Fermi surface is gapped in a structure where the Jones-zone boundaries touch the Fermi surface as much as possible.  Quasicrystals have such a structure where a combination of the strongest Bragg planes form an almost spherical Jones zone and by choosing the appropriate concentration and valence of the constituent elements the Fermi surface can be tailored to match the Jones zone.

\begin{figure}
\includegraphics[width=3.5in]{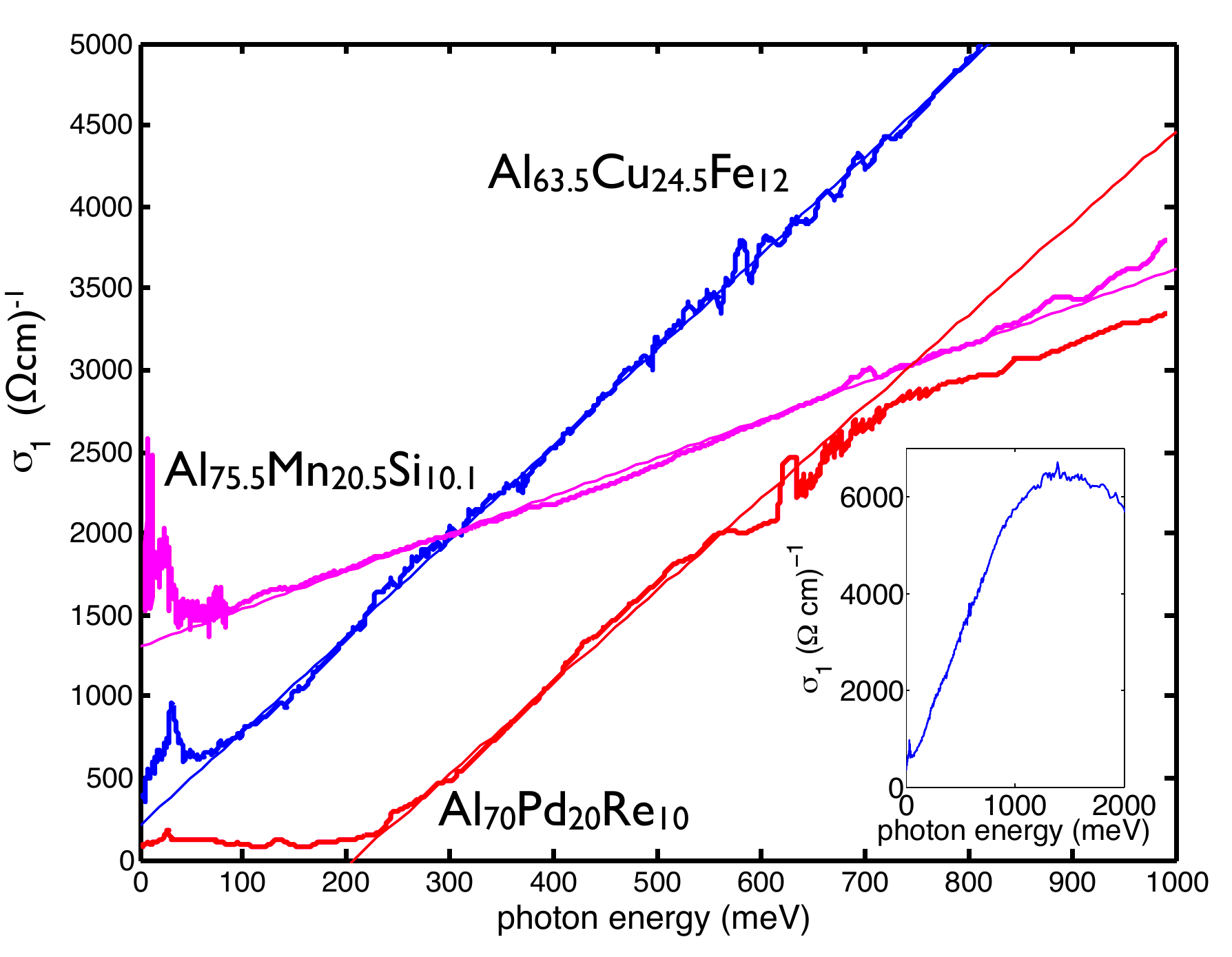}%
\caption{\label{ico} (color on line)The optical conductivity of three icosahedral quasi crystals measured at 300 K.  The solid lines are experimentally measured conductivities and the thin lines least squares fits of straight line segments to the data. All spectra show evidence of phonon lines below 100 meV but surprisingly as the overall low frequency conductivity increases the phonons get stronger.  Unlike conventional metals these materials show no evidence of a Drude peak from free electrons.  The linear interband conductivity shows differences too.  In AlCuFe  (Ref. 12) there is a single linear component intercepting near zero frequency while the AlPdRe material (Ref. 14) has a negative intercept that we interpret, see below, as a 3D Dirac point some 100 meV from the Fermi surface.  The AlMnSi material (Ref. 13) has a strong incoherent background that seems to have transferred spectral weight from the inter band absorption which has a lower slope than the other two.  We also note that the steeper slopes are very similar in AlCuFe and AlPdRe while the slope above 700 meV if AlPdRe matches that of AlMnSi.   The structure in the 600 meV region in all the spectra is due to instrumental noise.  The inset shows the AlCuFe conductivity up to 2 eV. }
\end{figure}

The optical conductivities of the three dimensional icosahedral quasi crystals and their periodic approximants  are quite remarkable. First they lack the Drude peak characteristic of free electrons but even more unusual is the frequency dependence of the conductivity \cite{homes91}. Figure 1 shows the conductivity spectra for three icosahedral quasicrystals. There are some striking common elements in these curves.  All start from a low conductivity at low frequency and rise linearly up to approximately 1.0 eV where there is a broad maximum.  There are some additional structures at low frequency which can be attributed to phonons\cite{homes91,wu93,basov94}. For example in Al$_{63.5}$Cu$_{24.5}$Fe$_{12}$ (AlCuFe) from Homes \etal\cite{homes91} there is a phonon contribution below 50 meV but no sign of a Drude peak.  In contrast, in pure aluminum there is a Drude peak with an amplitude of $\sigma(0) = 3.5\times10^5\ (\Omega$cm)$^{-1}$ and a width $\gamma = 82$ meV at room temperature arising from electron-phonon scattering.  A linear fit to the data between 200 and 800 meV (the thin line) extrapolates to zero frequency at a positive intercept of 350 $(\Omega$cm)$^{-1}$.  This is in qualitative agreement with the low dc resistivity becoming even lower as the sample quality improves.   On the whole, setting aside for the moment the phonon contribution and the weak dc component, the optical conductivity of this material has only one major component: a conductivity with a striking linear rise between 0 and 1 eV that eventually saturates at 1.4 eV as shown in the inset to Fig. 1. 

The Al$_{75.5}$Mn$_{20.5}$Si$_{4}$ (AlMnSi) curve in Fig. 1 is from Wu \etal\cite{wu93}.  Here the zero frequency intercept is much higher, $\sigma(0) = 1350\  (\Omega$cm)$^{-1}$; there is still a linear high frequency component but with a markedly lower slope.  It appears that spectral weight has been transferred from the interband absorption to a background component.  The third  curve shows  Al$_{70}$Pd$_{20}$Re$_{10}$ (AlPdRe) from Basov\etal\cite{basov94}.  This material seems to have three components to the conductivity.  At the lowest frequencies there is a weak, frequency independent component up 220 meV, followed by a linear rise up to 700 meV at which point there is a change of slope followed by another linear rise.  It is interesting to note that the slopes of the two linear components in (AlPdRe)  match closely the slopes of single lines in the other two compounds.

The sharp structure below 100 meV matches roughly the phonon density of states in metallic aluminium\cite{homes91}.  However the overall amplitudes of the phonon spectra increase as the background conductivity at low frequency {\it increases}.  This is just the opposite to what is naively expected for a metal: the higher the metallic conductivity, the stronger is the screening of the ionic charges by the electrons and the weaker is the phonon conductivity.

\begin{figure}
\includegraphics[width=3.5in]{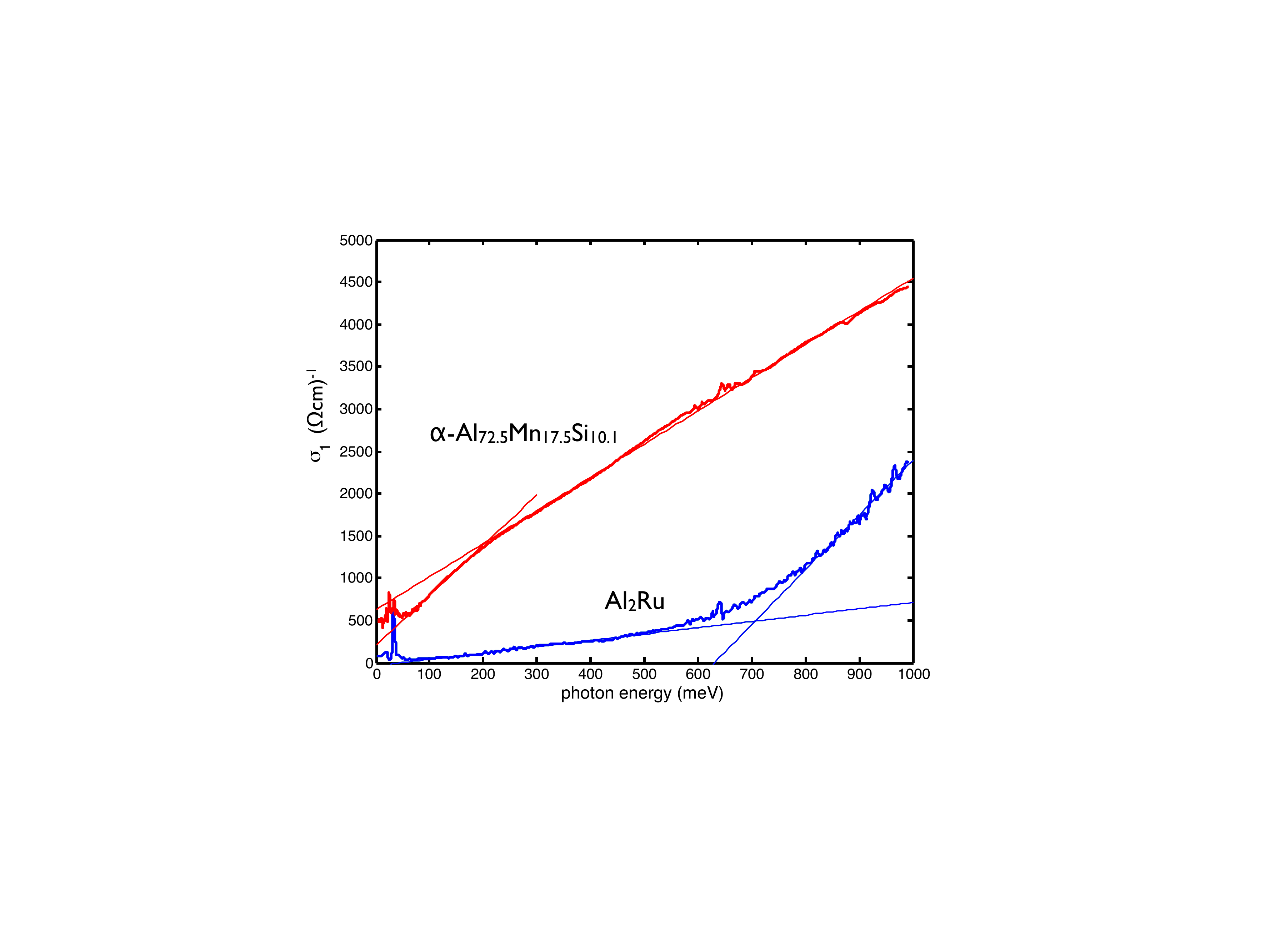}%
\caption{\label{approx} (color on line)The optical conductivity of two periodic approximants of quasicrystals at 300 K.  These materials resemble quasicrystals in their overall structure. There is an absence of a Drude peak and the conductivity is dominated by linear segments. Al$_2$Ru from Ref. 14 has two components, a very weak one that intercepts the conductivity axis near zero frequency and a stronger component that shows a gap on the order of 300 meV.  The $\alpha$-AlMnSi material from Ref. 13 has a slope that is intermediate between the large and small slopes of the icosahedral quasicrystals and a semiconductor-like gap at very low frequency.}
\end{figure}

Figure 2 shows the optical conductivity of two periodic approximants to quasicrystals.  They are qualitatively similar to the quasicrystals shown in Fig. 1. There are no Drude peaks and the spectrum consists of segments of straight lines.  There are however notable differences as well.  In particular the  $\alpha$-Al$_{72.5}$Mn$_{17.5}$Si $_{10.1}$ ($\alpha$-AlMnSi) from Wu \etal\cite{wu93} shows a low frequency downturn below 200 meV.  One can perhaps argue that this material has a semiconducting gap at very low frequency with a square root turn-on. However it is difficult to fit the overall curve with a parabola. Finally the Al$_2$Ru spectrum from Basov \etal\cite{basov94} also shows a two-component spectrum, a low-slope low-intercept part below 600 meV and a higher frequency component with an onset at 700 meV.

\begin{figure}
\includegraphics[width=3.5in]{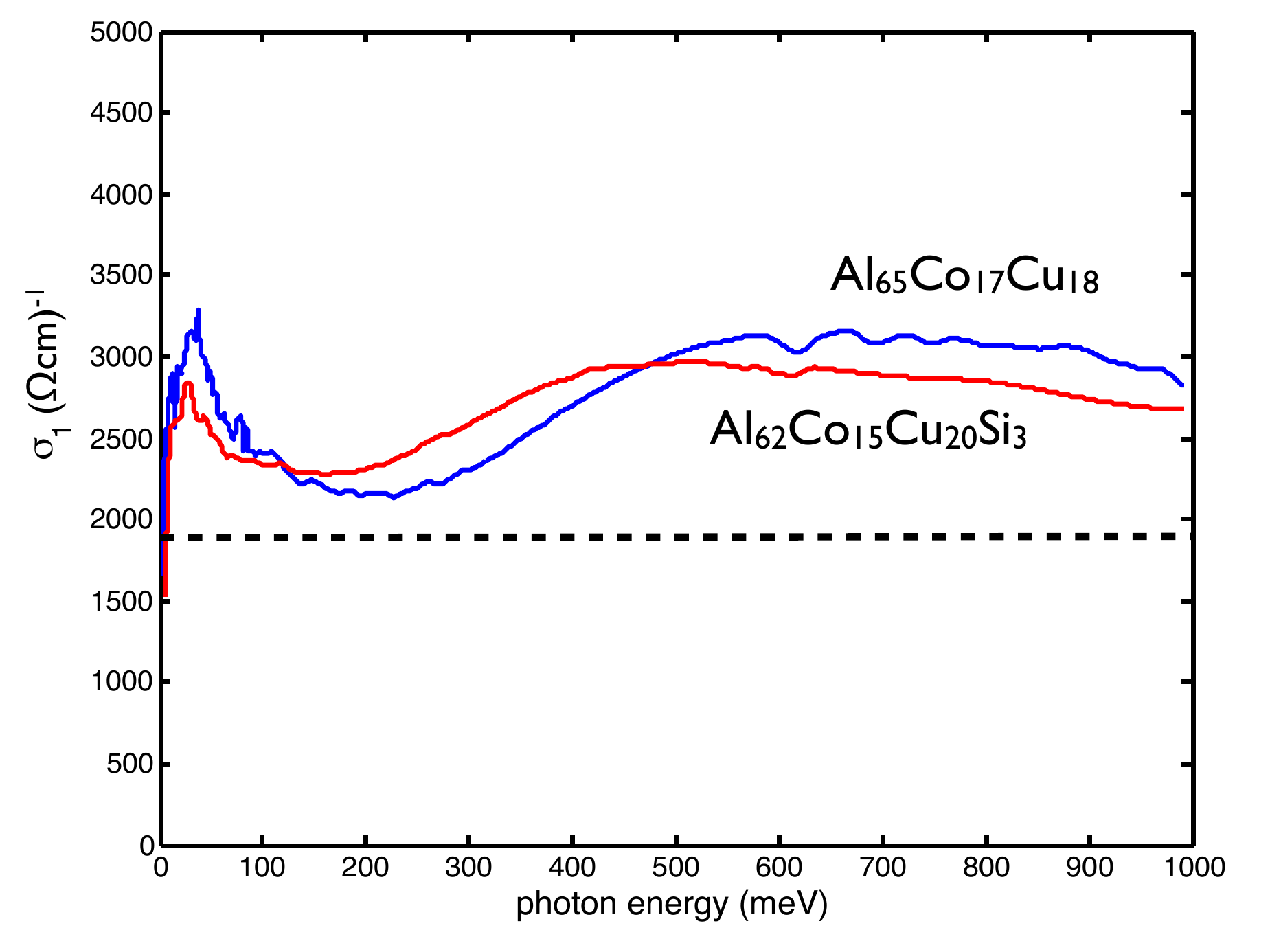}%
\caption{\label{dec} (color on line) Decagonal quasicrystals at 300 K from Ref. 15.  These curves show a conductivity that is, to a first approximation, frequency independent but has a slight downturn at low frequency that can be interpreted as a gap in a 2D Dirac scenario.  In addition to the phonons below 100 meV one can perhaps make a case for a weak Drude band with a damping of the order of 100 meV.  These materials show a clear Drude peak in the perpendicular periodic direction. The dashed line is  at $\sigma=G_0/c=1880 \ (\Omega$cm)$^{-1}$ where $G_0$ is the quantum conductance and c =4.12 \AA,   the lattice spacing .}
\end{figure}

We next turn to  the optical conductivity of decagonal quasicrystals.  These materials exhibit quasicrystalline periodicity in a plane  but have conventional periodic symmetry in the third direction.  Optical spectroscopy shows that the conductivity in the periodic direction is that of a typical metal with a well defined Drude peak\cite{basov94a} with a width $\hbar/\tau=50$ meV for Al$_{65}$Co$_{17}$Cu$_{18}$ and         $\hbar/\tau=160$ meV for Al$_{62}$Co$_{15}$Cu$_{20}$Si$_3$ but in the decagonal plane the conductivity, as shown in Fig. 3, is quite different, weakly frequency dependent up to 1 eV.  As in the icosahedral quasicrystals there is no metallic Drude peak in the icosahedral plane.  The magnitude of the weakly frequency dependent conductivity is fairly low, expressed as the conductance per plane $G/c$, it is fairly close to universal quantum conductance $G_0=2e^2/h=7.75\times10^{-5} \Omega^{-1}$, shown as a dashed line in Fig. 3.  Below 150 meV we see a phonon spectrum similar to what is seen in the 3D icosahedral systems.  There is a downturn in conductance between 500 and 200 meV that could be taken as evidence of a partial gap near zero frequency.  On the whole the spectra resemble those of the two dimensional graphene  both in frequency dependence and in the overall magnitude of the conductivity\cite{kuzmenko08}.   \\

\section{Theoretical background}

The linear frequency dependence of the optical conductivity is unusual and not easily understood in terms of what is observed in conventional materials.  The optical conductivity of metals is, at low frequency, dominated by the Drude conductivity of the free electrons that collide with static defects with an average collision time $\tau$.  Insulators, on the other hand are characterized by a low frequency region of zero conductivity with a sharp onset at the band gap  frequency where inter band transitions  first set in.

Disordered systems, for example as measured by Th\`eye \etal\cite{theye85}, show a Drude-like optical conductivity with a very high scattering rate of $\hbar/\tau$ of the order of 0.6 eV with a good agreement between optical and dc measurement, a signature of a Drude behavior extending to low frequencies. Attempts to describe the optical data of quasicrystals in terms of a partially gapped Fermi surface have been made\cite{burkov92} but with limited success. For example, in metallic aluminum the inter band absorption rises quadratically at low frequencies\cite{ashcroft71}.  It is difficult to model a linear conductivity down to zero frequency unless one assumes a Fermi surface with zero diameter pocket on the Fermi surface and a massless Dirac spectrum.  
In what follows we will make this assumption and calculate the optical conductivity of a system of massless 3-D fermions.

Here we follow the nearly free electron approach to the electronic structure of quasicrystals taken by Burkov {\it et al.}\cite{burkov92}. In this approach the electron momentum is a good quantum number and the the Fermi surface is taken to be in good contact with the several prominent Bragg planes seen in crystallography. Other subdominant Bragg planes do not play an important role.  In this picture these materials are semi metals and their low conductivity is not due to a short mean free path but is rather due to a low concentration of itinerant carriers. Bands could cross at momentum points even when not necessitated by crystal symmetry as described in the early works of Herring \cite{herring37} and can generate Dirac points without a need for strong spin orbit coupling. Because we do not have inversion symmetry the degeneracy of the Dirac points is expected to be lifted providing two stable Weyl points separated in energy which could be near the Fermi energy.\cite{vazefeh13}  Any other power law for the electron dispersion, say $\epsilon(\bf k) \propto {|k|} ^{\it z}$ would give rise to an inter band contribution to the dynamic optical conductivity $\sigma_1(\omega) \propto |\omega|^{\frac{D-2}{z}}  $where $D$ is the dimensionality of the space (here $D=3$)\cite{bacsi13}. The observation of a linear in $\omega$ law necessitates  $\epsilon(\bf k) \propto |k|$. Thus we adopt a Weyl semimetal model.
 
The Hamiltonian for a 3-D Weyl or massless Dirac fermions with fixed handedness 
which is described by a two-component spinor can be written as
\begin{equation}
H=\hbar v_F \mbox{\boldmath $\tau \cdot k$} - \mu \tau_0,
\end{equation}
where $v_F$ is the Fermi velocity, and $\tau_i$ with $i=x,y,z$ are the Pauli matrices which take care of the pseudospin degree of freedom, $\tau_0$ is the unit $2\times 2$ matrix, ${\bf k}$ is the 3-D wave-vector, and $\mu$
is the chemical potential.  The energies  of the quasiparticle excitations are linear in momentum,
$\xi_{\bf k}= \pm\hbar v_F|{\bf k}| - \mu$, i.e. relativistic with $v_F$ replacing the velocity of light.
While quasicrystals do not have long range crystalline order, crystallographic analysis reveals that some 
Bragg planes remain.  As we do not see strong evidence of gaps we suggest the that the linear dispersion originates from Weyl points.
% We will use Weyl from here on.

To evaluate the diagonal optical conductivity we use the Kubo formula,
\begin{equation}\label{Kubo}
{\Re e\sigma(\Omega)}= - \frac{\hbar {\Im m\Pi^{R}_{ii}(\Omega+i0)}}{3\Omega}, 
\end{equation}
where $\Pi^{R}_{ij}(\Omega+i0)$ is  the retarded current-current correlation
function, $\Omega$ is the energy of photon,  and the sum over the repeated index $i$ is implied. 
In the lowest approximation
\begin{equation}
\label{Pi-via-SF}
\begin{split}
\Pi_{i j}^{R}(\Omega+i0)=& {e^2v_F^2}\int\limits_{-\infty}^\infty d\omega
d\omega^\prime\frac{n_F(\omega)-n_F(\omega^\prime)}
{\omega-\omega^\prime-\Omega-i0} \\
& \times \int\frac{d^3 k}{(2\pi)^3}{\rm
tr}\left[\tau_i A(\omega,\mathbf{k})\tau_j
A(\omega^\prime,\mathbf{k})\right],
\end{split}
\end{equation}
where $e$ is the electron charge, $n_F(\omega)=1/(\exp((\omega-\mu)/T)+1)$ is the Fermi distribution, and
$\mbox{tr}$ includes the sum over $N_W$ Weyl points. Since the spin degeneracy is assumed 
to be lifted, the sum over spin degree of freedom is included in the number $N_W$.
In Eq.~(\ref{Pi-via-SF}) the spectral function  is given by
\begin{equation}
\label{SF}
\begin{split}
A(\omega,\mathbf{k}) &=  \delta(\omega -\epsilon_\mathbf{k}) \frac{1}{2}
\left(\tau_0 +  \frac{\hbar v_F \mbox{\boldmath $\tau \cdot k$} }{\epsilon_\mathbf{k}} \right) \\
&
+\delta(\omega + \epsilon_\mathbf{k}) \frac{1}{2}
\left(\tau_0  - \frac{\hbar v_F \mbox{\boldmath $\tau \cdot k$} }{\epsilon_\mathbf{k} } \right),
\end{split}
\end{equation}
where $\epsilon_\mathbf{k} = \hbar v_F |\mathbf{k}|$. One can easily see from Eq.~(\ref{SF}) that the spectral 
function of the Weyl fermions selects that the electron (hole)-like excitations to posses 
the positive (negative) helicity. The resulting optical conductivity consists of two pieces, 
viz. intraband and interband.
To simplify our presentation Eq.~(\ref{SF}) is written assuming that there is no
broadening caused by interactions and scattering from impurities.
It is well known, however, that to reproduce correctly the Drude (intraband) part of the conductivity 
from the Kubo formula (\ref{Kubo}), one has to include the disorder, see Bradlyn {\it et al.}\cite{bradlyn12}
When this is done, one can take the limit $\tau \to \infty$ at the end of the calculation.
Then for zero temperature, $T=0$ the resulting intraband piece takes the form
\begin{equation}
{\Re e\sigma^{\mathrm{intra}}}(\Omega)= \frac{ N_W e^2\mu^2}{ h \hbar v_F} {{1} \over {3}}\delta(\Omega),
\end{equation}
while the interband which onsets only above $2|\mu|$ is given by
\begin{equation}
{\Re e\sigma^{inter}(\Omega)}={{N_W e^2 |\Omega|} \over { h{\hbar v_F}}}{{1} \over {12}}\Theta({|\Omega|-2|\mu |)}
\end{equation}
and is linear in energy $\Omega$. It is clear from these expressions that for finite chemical potential 
$\mu$ the optical spectral weight in the inter band piece lost below $2|\mu|$ gets transferred to 
the intraband Drude piece. When the chemical potential
$\mu =0$, the intraband piece vanishes and the remaining interband part reduces to the expression given 
in.\cite{hosur12}

\section{Discussion}
Eq. (6) predicts, for the case where the Weyl point lies on the Fermi surface,  an interband spectrum consisting of a linear rise in the optical conductivity with a zero intercept.  The slope is proportional to the number of Weyl points divided by the Fermi velocity.  If the Weyl point is not on the Fermi surface the straight line for the conductivity would still extrapolate to the origin.  However it would terminate at $\Omega=2|\mu|$ where it would abruptly fall to zero regardless of the sign of $\mu$.

None of the quasicrystal spectra in Fig. 1 meet these strict criteria but all display some Dirac like features.  AlPdRe has a linear portion extending from 210 meV to 700 meV which would however intercept the x-axis at a finite photon energy of $\Omega=210$ meV rather than go through the origin. This rigid shift of the curve to the right by 210 meV would arise in a Dirac like model if the common Dirac points of the conduction and valence bands were split by 210 meV for a not yet known reason.

Added to this is a weak frequency independent constant background of about 150 $(\Omega$cm)$^{-1}$.  There is a break in slope at 700 meV.   It can be argued that the AlCuFe material presents an even simpler confirmation of Eq. (6) with $\mu=0$ provided we neglect the small {\it negative} intercept on the frequency axis.  This negative intercept is inconsistent with Eq. (6). To account for the negative intercept we will assume that there is a second {\it frequency independent} channel of conductivity in the region from zero to 1000 meV.  We will call this constant background the free electron contribution since it is finite at zero frequency.  The slopes of the linear portion in the two materials are similar.  The third quasicrystal AlMnSi also has the linear conductivity predicted by Eq. (6) but the slope is considerably smaller and the frequency independent channel is much stronger. Also the phonon spectrum is much stronger in AlMnSi.

We would expect the total spectral weight to be conserved and it does appear to be the case for AlCuFe and AlMnSi where the inter band linear contribution of AlCuFe is much larger than in AlMnSi but compensated by the larger intraband contribution in the latter material. The total areas under the curves in Fig. 1 become equal at approximately 600 meV at which point the AlCuFe partial spectral weight increases more rapidly.   This compensation does not take place in AlPdRe where the inter band contribution is much smaller than in the other two materials, due mostly to the gap at 210 meV, but is not made up by the very small free carrier contribution.

It is instructive to compare the magnitude of the measured conductivity with the value predicted by Eq. (6).  We will use the AlCuFe system as an example where the conductivity rises linearly with a slope of 5750 $(\Omega$cm)$^{-1}$/eV yielding us the ratio of the conductivity $\sigma$ to the energy $\Omega$.  If we assume that the Weyl points are located on the faces of the icosahedron as shown in Fig. 4, we get 20 such points and with a spin degeneracy of two $N_W=40$.  We can now calculate the only remaining unknown, the Fermi velocity:
\begin{equation}
v_F = {N_W \over 4}G_0{\omega \over \sigma(\omega)} = 4.3 \times 10^7\  {\rm cm/sec}
\end{equation}
where $G_0 = 2e^2/h= 7.748\times 10^{-5}\  \Omega^{-1}$ is the quantum conductance. A rough estimate the Fermi energy is the energy of a free electron with a velocity $v_F$, $E=v_F^2/2m_e =  0.5$ eV. This is a reasonable value if states below the Fermi surface follow a free electron parabola as shown in the inset of Fig. 1, from the the data in Homes {\it et al.}\cite{homes91}.  This  good agreement is certainly accidental in view of our naive estimate of the Fermi velocity and the assumption of the location of the Weyl points on the icosahedron.  The data of Nayak {\it et al.}\cite{nayak12} show a kink in the density of states at Å 0.5 eV yielding the same Fermi velocity by the same argument. 
Another rougher estimate can be made from ARPES data of Rotenberg {\it et al.}\cite{rotenberg00} on the decagonal quasicrystal AlNiCo where there is band dispersing to flatten out at 2 eV below the Fermi surface yielding a higher Fermi velocity.  

\begin{figure}
\includegraphics[width=3.0in]{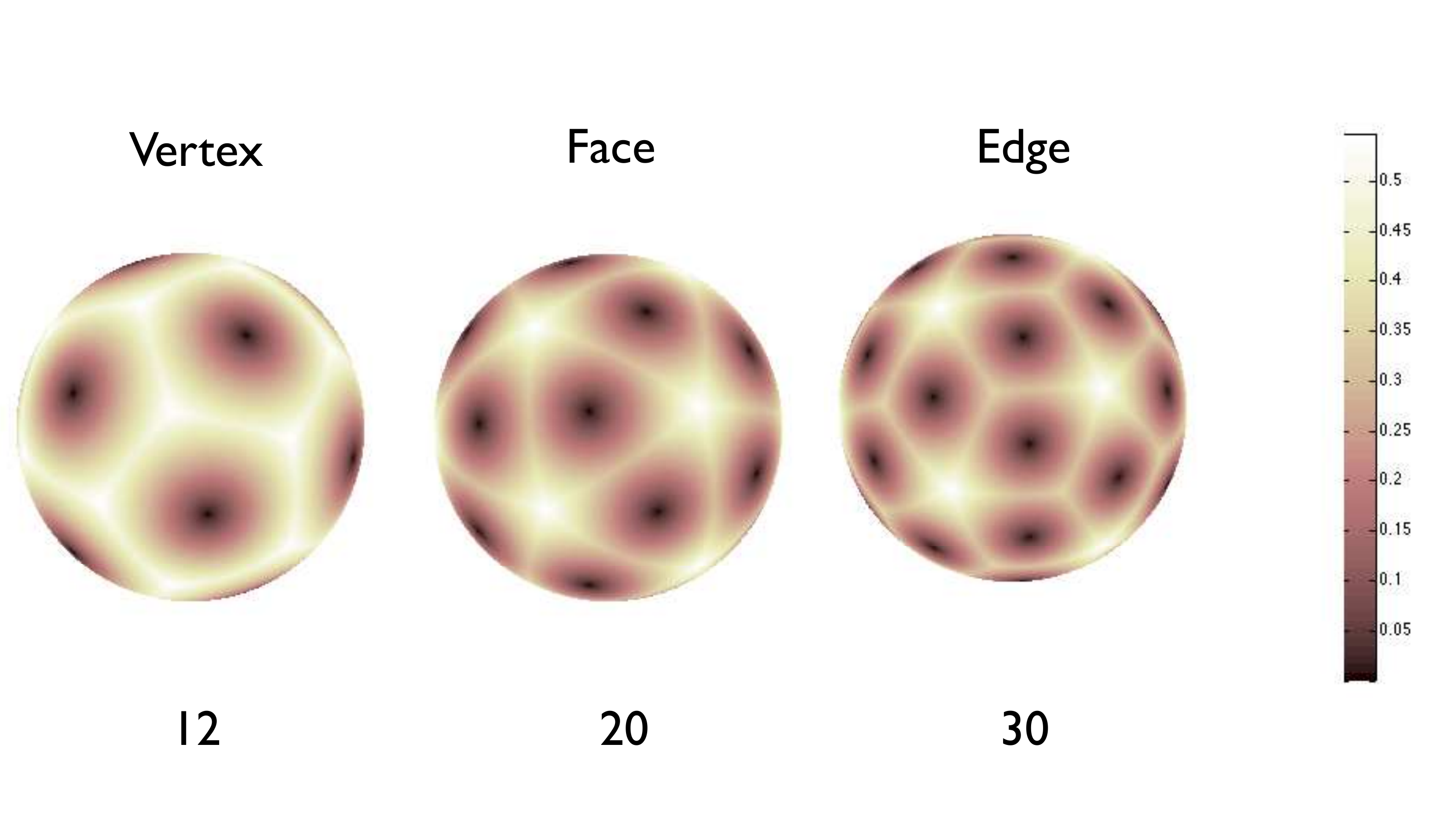}%
\caption{\label{icosahedron} (color on line) Possible band dispersions radiating from nodal points (marked in black) at various symmetry points of an icosahedron from left to right, 12 vertices , 20 faces and 30 edges. The energy scale, shown on the right, starts at  $E=0$ (black)  at the Dirac points to a maximum, (white) at $E=0.5$ eV. }
\end{figure}

The periodic approximants in Fig.2 show similar linear conductivities rising from near zero up to 1000 meV.  The $\alpha -$AlMnSi shows a break of slope at 210 meV and one might be tempted to fit the curve with a square root onset near zero frequency but clearly a pair of linear fits is better. The Al$_2$Ru spectrum is different from all the others in that it has two well defined components.  The first is a linear one that intercepts at zero frequency, according to Eq. (6) this component has a Weyl point on the Fermi surface but a very small slope implying a very high Fermi velocity, or alternatively few Weyl points. A second component exhibiting a rigid displacement of 630 meV seen for AlPdRe in Fig. 1. The slope of this component is similar to what is seen in other materials.

The spectra of the two decagonal quasicrystals shown in Fig. 3, to a first approximation, are frequency independent, constant with a conductivity of 2500 ($\Omega$cm)$^{-1}$ between zero and 1000 meV.   The dashed line is drawn at $\sigma_1 = G_0/c =1880 \   (\Omega$cm)$^{-1}$ and  c=4.12 \AA  \ is the c axis lattice spacing.

It is useful to compare our decagonal quasi crystal conductivity to that of graphite measured by Kuzmenko \etal \cite{kuzmenko08}.  These authors find that the optical sheet conductance of graphite per graphite layer is very close to the theoretically expected sheet conductance of a monolayer of graphene.  Our value is higher than this by a factor of about 1.5.  In graphite the measured conductance also exceeded the universal conductance in the energy range 0.6 eV to 1.2 eV which the authors  attributed to contributions from interlayer hopping.  This may be the case here too since the c-axis  conductance of our sample is quite high.

Taking this comparison further one could interpret the structure below 500 meV in Fig. 3 as due to a Dirac point some 200 meV below the Fermi surface with a transfer of spectral weight to a weak Drude peak with a width of the order of 100 meV. The Drude width in the periodic direction is 160 meV for AlCoCuSi and 50 meV for AlCoCu in rough agreement with this picture. 

Oxides with large spin-orbit interaction such as the pyrochlore iridates R$_2$Ir$_2$O$_7$ where R is Yttrium or a rare earth have been proposed as model systems for TSM (topological semi metals) \cite{wan11,yanagishima01} as well as osmium compounds such as CaOs$_2$O$_4$ \cite{wan12}.   Another system where a Dirac-like spectrum has been predicted is the 2D organic  material $\alpha$-(BEDT - TTF)$_2$I$_2$ \cite{goerbig08}. Here we  add quasicrystals and their approximants to the list. To show conclusively that quasicrystals and their approximants are Weyl semi metals it is important to perform experiments in high magnetic fields.  For the quasicrystals these will be challenging since it is not clear that coherent orbits can be generated in available laboratory magnetic fields.  However, approximants such as Al$_2$Ru are candidates for such experiments if pure single crystals are available. Because of the Dirac nature of their dispersion curves the conductivity in high magnetic field should show a structure each time a Landau level is crossed. The prediction is that the position in energy of these peaks should vary like the square root of the magnetic field B.  Also, the optical conductivity of any new candidate materials should be collected since it provides clear signatures of relativistic dispersions in 3D: a linear frequency dependence of the optical conductivity.  As an example  a recent study by Ueda \etal\cite{ueda12} on Nd$_2$(Ir$_{1-x}$Ru$_x$)$_2$O$_7$ with x=0.02,  a proposed Weyl semi metal, found some evidence of a Dirac-like spectrum below 40 meV.

% figures should be put into the text as floats.
% Use the graphics or graphicx packages (distributed with LaTeX2e)
% and the \includegraphics macro defined in those packages.
% See the LaTeX Graphics Companion by Michel Goosens, Sebastian Rahtz,
% entire page. There is no need to do explicit centering.

%\appendix
%\section{}

\begin{acknowledgments}

The authors would like to thank Dimitri Abanin, Anton Burkov, Sung-Sik Lee, Marcel Franz, Xiao-Gang Wen and Elizabeth Nicol for helpful discussions.  
This work was supported in part by the Canadian Institute for Advanced Research and the Natural Science and Engineering Research Council of Canada.  The work at UCSD was supported by DOE-BES. The work of S.G.Sh. was supported by the European FP7 program, Grant
No. SIMTECH 246937, a collaborative grant from the Swedish Institute, and
by the Grant STCU \# 5716-2 "Development of Graphene Technologies and Investigation of 
Graphene-based Nanostructures for Nanoelectronics and Optoelectronics.

\end{acknowledgments}
\eject

\end{document}